\documentclass[twocolumn]{autart}

\usepackage{amssymb,amsmath,stackrel,amsopn,latexsym}
\usepackage{graphicx}
\usepackage{amsfonts}
\usepackage[authoryear,round]{natbib} 
\usepackage{color}

\usepackage{enumitem}

\graphicspath{{img/}}

\newcommand*{\red}{\textcolor{black}}
\newcommand*{\blue}{\textcolor{black}}

\newcommand{\R}{\mathbb{R}}
\newcommand{\E}{\mathbb{E}}
\newcommand{\M}{\mathcal{M}}

\newcommand{\Sy}{\mathcal{S}}
\newcommand{\Eb}{\bar{\mathbb{E}}}

\newtheorem{definition}{Definition}
\newtheorem{proposition}{Proposition}
\newtheorem{theorem}{Theorem}
\newtheorem{example}{Example}

\newtheorem{remark}{Remark}
\newtheorem{lemma}{Lemma}
\newtheorem{assumption}{Assumption}

\makeatletter
\renewcommand*{\@opargbegintheorem}[3]{\trivlist
  \item[\hskip \labelsep{\bfseries #1\ #2}] \textbf{(#3)}\ \itshape}
\makeatother

\begin{document}

\begin{frontmatter}
\title{Prediction error identification of linear dynamic networks \\ with rank-reduced noise \thanksref{footnoteinfo}
}

\thanks[footnoteinfo]{Paper submitted for publication in {\it Automatica}, 17 november 2017. Corrected 8 December 2017. Revised for Automatica, 4 April 2018.
This project has received funding from the European Research Council (ERC), Advanced Research Grant SYSDYNET, under the European Union's Horizon 2020 research and innovation programme (grant agreement No 694504).}

\author[First]{Harm H.M. Weerts},
\author[First]{Paul M.J. Van den Hof} and
\author[Second]{Arne G. Dankers}

\address[First]{Control Systems Group, Department of Electrical Engineering, Eindhoven University of Technology, The Netherlands (email: h.h.m.weerts@tue.nl, p.m.j.vandenhof@tue.nl)}
\address[Second]{Department of Electrical Engineering, University of Calgary, Canada (email: adankers@hifieng.com)}


\begin{abstract}
Dynamic networks are interconnected dynamic systems with measured node signals and dynamic modules reflecting the links between the nodes. We address the problem of \red{identifying a dynamic network with known topology, on the basis of measured signals}, for the situation of additive process noise on the node signals that is spatially correlated and that is allowed to have a spectral density that is singular. A prediction error approach is followed in which all node signals in the network are \red{jointly} predicted. \red{The resulting joint-direct identification method, generalizes the classical direct method for closed-loop identification to handle situations of mutually correlated noise on inputs and outputs. When applied to general dynamic networks with rank-reduced noise, it appears that }
the natural identification criterion becomes a weighted LS criterion that is subject to a constraint. This constrained criterion is shown to lead to maximum likelihood estimates of the dynamic network and therefore to minimum variance properties, reaching the Cram\'{e}r-Rao lower bound in the case of Gaussian noise.
\end{abstract}
\end{frontmatter}

\section{Introduction}
\label{Sec:Introduction}
It is becoming more common to model complex dynamic systems as networks of interconnected dynamic modules, or \emph{dynamic networks}.
Data-driven modeling, or \emph{identification}, of modules in these dynamic networks is then a natural problem to address.
Applications range over many fields, for example identification of dynamics that connect different (MPC) control loops in industrial process control \citep{gudi2006identification,VandenHof&etal:CACE17},
identification of biochemical networks \citep{Yuan2011},
modeling of the dynamic behavior of a ship as a dynamic network \citep{linder2017},
and modeling of stock prices in financial markets as a dynamic network \citep{Materassi2010}.

Various approaches have been developed for identification of dynamic networks, roughly divided into three categories.
The first approach considers the identification of a single dynamic module in the dynamic network \red{in the situation that the interconnection structure, or topology, of the network is known.}
The second \red{approach} focusses on identification of the full network dynamics \red{for a given topology}, and the last category deals with the identification of the topology \red{(and dynamics)} of the network.
For identification of single modules, authors have used e.g. Wiener filters \citep{MaterassiSalapaka2012}, while parametric transfer functions can be estimated in a prediction error setting  \citep{VandenHof&Dankers&etal:13,gevers2015identification,dankers2015errors,dankers2016identification,linder2016identification}.
Identification of the full network dynamics can be done by modeling the network as a state-space system \citep{haber2014subspace}, or as a network of transfer function modules \citep{weerts2016cdc}. Identifiability properties related to this problem are addressed in \cite{Goncalves08,adebayo2012dynamical,weerts_etal_2015,weerts2016identifiability2,GEVERS2017ifac}.
Some different methods for topology detection can be found in literature, for example with a Bayesian approach \citep{ChiusoPAuto2012}, with a compressed sensing approach \citep{hayden2016sparse}, \red{and based on one-step ahead prediction using Wiener filters \citep{MaterassiSalapaka2012}}.

\begin{figure}[b]
	\centering
	\includegraphics[width=\columnwidth]{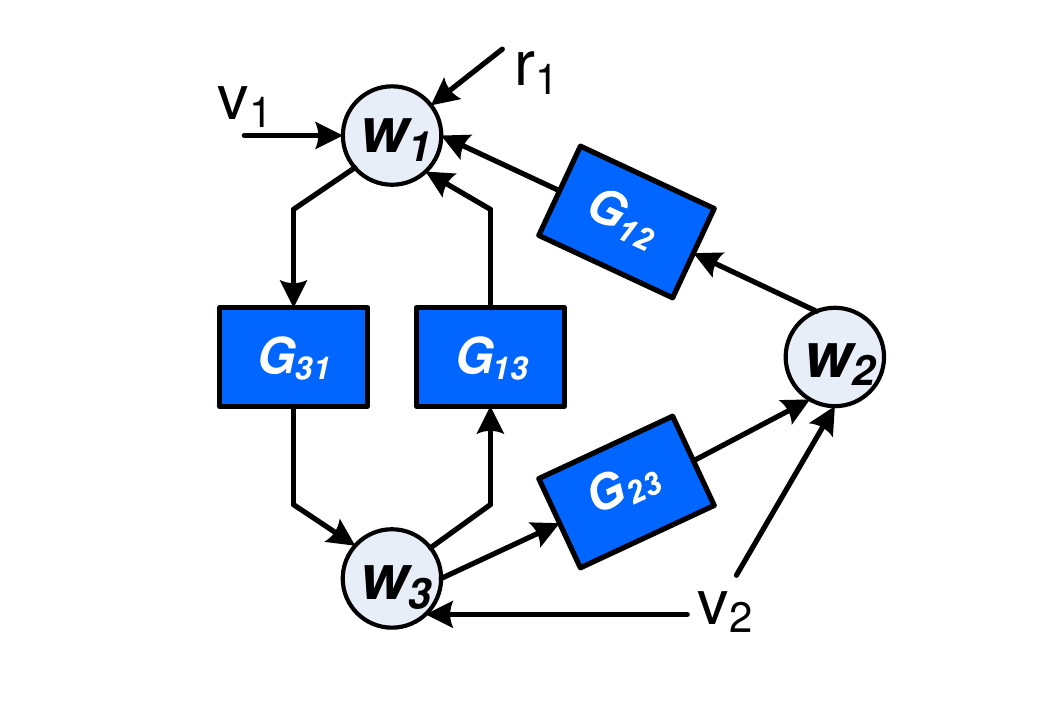}
	\caption{Example of a network with rank-reduced noise. Node signals are $w_i$, being the outputs of the (circular) summation points, interconnected by modules $G_{ij}$ and perturbed by non-measured disturbance signals $v_i$. Signals $r_i$ are excitation signals available to the user.}
	\label{fig:example_sys}
\end{figure}
In this paper we consider networks that consist of measured node signals, which are interconnected by linear dynamic modules, as depicted in Figure \ref{fig:example_sys}, and in line with the setup as defined in \cite{VandenHof&Dankers&etal:13}.
We will address the problem of \red{identifying, on the basis of measured time signals, the dynamics of all modules in a network, of which the topology is known. And we will do so} under conditions on the disturbance signals $v_i$ that are more general than typically considered. While in the current literature it is usually assumed that every node signal in the network has a non-zero process noise $v_i$ that is uncorrelated to all other noises, \red{i.e. for the vector noise process $v$ it holds that $\Phi_v(\omega)$ is diagonal, we will address two steps of generalization:
\begin{itemize}
\item We will allow noise signals on the different node signals to be spatially correlated, i.e. $\Phi_v(\omega)$ is not necessarily diagonal, and
\item We will allow $\Phi_v(\omega)$ to be singular, implying that node signals can be noise-free, or that disturbances are exactly related with each other through a linear filter.
\end{itemize}}
\red{Concerning the first step, it has to be remarked that this situation includes the handling of confounding variables, i.e. unmeasured variables that affect both inputs and outputs of an estimation problem. This notion is widely used in statistical estimation problems in networks and is also used in network identification problems, \cite{dankers2016identification}. The relation between confounding variables and correlated disturbances has been explained in \citep{VandenHof&etal:CDC17}}. \\
\red{Concerning the second step,} note that modules in a network can also be implemented controllers, and controller outputs can be noise-free, as e.g. typically considered in a classical closed-loop identification problem \citep{ljung:99}. In this case there is no process noise on a particular node signal. \red{Alternatively, strong correlations between disturbance signals can occur} e.g. if the network is a spatially distributed system affected by global disturbances, like a wind gust affecting wind turbines in a wind park. \red{A deterministic relation between disturbance signals (like e.g. a delay) will cause the full disturbance spectrum to loose its full rank.}
%
\red{A situation of loss of full rank} is depicted in Figure \ref{fig:example_sys} where the process noises on nodes $2$ and $3$ are the same (perfect correlation).
When identifying the full network dynamics, \red{aiming not only at consistency of the module estimates, but also at minimum variance results, that fact that disturbances are correlated will result in the situation that the identification problem can not be decomposed into separate multi-input single-output problems. The fact that the noise process is allowed to be} rank-reduced causes some fundamental issues that need to be addressed in the \red{prediction error identification setting.}

Identification in the situation of rank-reduced noise is a topic that has not been widely addressed in the prediction error identification literature. Dynamic factor models have been developed in \cite{Deistler2015,Fels_diss} to deal with rank-reduced noise.
Maximum likelihood estimates with rank-reduced noise have been obtained for vector autoregressive systems \citep{Kolbl_diss} and linear regression \citep{srivastava2002regression}.
In a prediction error setting, the property of \textit{network identifiability} has been defined in \cite{weerts_etal_2015,weerts2016identifiability2}, covering also the situation of rank-reduced noise, while predictor models have been analyzed for the situation of noise-free nodes in \cite{weerts2016identifiability} and for general rank-reduced noise in \cite{weerts2017ifacwc}. In \cite{weerts2017ifacwc} this has been extended with a first analysis of consistency of network dynamics estimates, leading to the use of weighted and constrained least-squares identification criteria, based on the preliminary work of \cite{VandenHof&etal:ACC17} where an open-loop one-input two-output situation with rank-reduced output noise was considered.

In this paper we are going beyond the consistency question, by including an analysis of the asymptotic variance of the prediction error method, and by developing the maximum likelihood estimator and the Cram\'{e}r-Rao lower bound on the variance, for the situation of \red{correlated and} rank-reduced noise, \red{while addressing networks with strictly proper modules.}
%
This paper builds on and further extends the preliminary results of \cite{weerts2017ifacwc}.

First a definition of the dynamic network setup and the rank-reduced noise process is given in Section \ref{Sec:NetDef}.
Then, in Section \ref{Sec:predictor}, the prediction error identification setup is presented and a least squares identification criterion is shown to provide consistent estimates.
In Section \ref{Sec:CLS} the dependencies in the noise process are explicitly used to construct a constrained least squares identification criterion that is shown to lead to a maximum likelihood estimate under some conditions.
An analysis of the asymptotic variance of the estimates is made in Section \ref{sect:var}, where the variance expressions are related to the Cram\'{e}r-Rao lower bound. Finally in Section \ref{Sec:simulation} the theoretical results are illustrated in a numerical simulation example.

\section{Dynamic network definition}
\label{Sec:NetDef}
Following the basic setup of \cite{VandenHof&Dankers&etal:13}, a dynamic network is built up out of $L$ scalar \emph{internal variables} or \emph{nodes} $w_j$, $j
= 1, \ldots, L$, and $K$ \emph{external variables} $r_k$, $k=1,\cdots K$.
Each internal variable is described as:
\begin{align}
w_j(t) = \sum_{\stackrel{l=1}{l\neq j}}^L
G_{jl}^0(q)w_l(t) + \sum_{k=1}^K
R_{jk}^0(q)r_k(t) + v_j(t)
\label{eq:netw_def}
\end{align}
where $q^{-1}$ is the delay operator, i.e. $q^{-1}w_j(t) = w_j(t-1)$;
\begin{itemize}[leftmargin=*]
	\item $G_{jl}^0$ are strictly proper rational transfer functions, and the single transfers $G_{jl}^0$ are referred to as {\it modules} in the network.
	\item $r_k$ are \emph{external variables} that can directly be manipulated by the user,
	and $R_{jk}^0$ are proper rational transfer functions;
	\item $v_j$ is \emph{process noise}, where the vector process $v=[v_1 \cdots v_L]^T$ is modelled as a stationary stochastic process with rational spectral density, such that there exists a $p$-dimensional white noise process $e:= [e_1 \cdots e_p]^T$, $p \leq L$, with covariance matrix $\Lambda^0>0$ such that
\[ v(t) = H^0(q)e(t), \]
\end{itemize}
with $H^0(q)$ a proper rational transfer function.

When combining the $L$ node signals we arrive at the full network expression
\begin{align*}
\begin{bmatrix}  \! w_1 \!  \\[7pt] \! w_2 \!  \\[1pt]  \! \vdots \! \\[1pt] \! w_L \!  \end{bmatrix} \!\!\! = \!\!\!
\begin{bmatrix}
0 &\! G_{12}^0 \!& \! \cdots \! &\!\! G_{1L}^0 \!\\
\! G_{21}^0 \!& 0 & \! \ddots \! &\!\!  \vdots \!\\
\vdots &\! \ddots \!& \! \ddots \! &\!\! G_{L-1 \ L}^0 \!\\
\! G_{L1}^0 \!&\! \cdots \!& \!\! G_{L \ L-1}^0 \!\! &\!\! 0
\end{bmatrix} \!\!\!\!
\begin{bmatrix} \! w_1 \!\\[7pt]  \! w_2 \!\\[1pt] \! \vdots \!\\[1pt] \! w_L \! \end{bmatrix} \!\!\!
+ \!\! R^0 \! (q) \!\!\!
\begin{bmatrix} \! r_1 \!\\[7pt] \! r_2 \!\\[1pt] \! \vdots \!\\[1pt]  \! r_{K} \!\end{bmatrix}
\!\!\!+\!\!
H^0 \! (q) \!\!\! \begin{bmatrix}\! e_1 \!\\[7pt] \! e_2 \!\\[1pt] \! \vdots \!\\[1pt] \! e_p\!\end{bmatrix} \!\!\!
\end{align*}
Using obvious notation this results in the matrix equation:
\begin{align} \label{eq.dgsMatrix}
w = G^0 w + R^0 r + H^0 e.
\end{align}
The network transfer function that maps the external signals $r$ and $e$ into the node signals $w$ is denoted by
\begin{equation}
\label{eq:T0}
T^0(q) := \begin{bmatrix} T_{wr}^0(q) & T_{we}^0(q) \end{bmatrix},
\end{equation}
where
\begin{align}
	T_{wr}^0(q) &:=  \left( I-G^0(q)\right)^{-1} R^0(q),
	\\
	T_{we}^0(q) &:=  \left( I-G^0(q)\right)^{-1} H^0(q). \label{eq:twe}
\end{align}
\red{Note that by choosing $G^0 = 0$ the network reduces to a multivariable open-loop system with inputs $r$ and outputs $w$.}

The noise component $\bar v(t)$ is defined according to $\bar v(t) := w(t) - T_{wr}^0(q)r(t)$ and satisfies
\begin{equation}
	\bar v(t) = T_{we}^0(q) e(t),
\end{equation}
with power spectral density
\begin{equation} \label{eq:vbar}
	\Phi_{\bar v}(\omega) := T_{we}^{0}(e^{i\omega}) \Lambda^0 T_{we}^{0,T}(e^{-i\omega}).
\end{equation}
This power spectral density can be determined using
\begin{equation}
	\Phi_{\bar v}(\omega) = \Phi_w(\omega) - T_{wr}^{0}(e^{i\omega}) \Phi_r(\omega) T_{wr}^{0,T}(e^{-i\omega}),
\end{equation}
where $\Phi_w$ and $\Phi_r$ are the power spectral densities of $w$ and $r$ respectively.

The noise model $H^0$ requires some further specification. For $p=L$, referred to as the \emph{full-rank} noise case, $H^0$ is square, stable, monic and minimum-phase. The situation $p < L$ will be referred to as the \emph{singular} or \emph{rank-reduced} noise case.\\
For notational simplicity and without loss of generality the following assumption will be made.

\begin{assumption} \label{asspi}
The $L$ node signals $w_j$, $j=1, \cdots L$ are ordered in such a way that
$ \left [ v_1  \cdots  v_p  \right ]^T$ is a full rank noise process.
\hfill $\Box$
\end{assumption}
How this assumption can be dealt with in actual identification will be discussed later on in Remark \ref{rem:order}.

The ordering of the noise signals gives rise to a representation for $H^0$ that satisfies
\begin{equation}
\label{eqH}
H^0(q) = \begin{bmatrix} H_a^0 \\ H_b^0 \end{bmatrix}
\end{equation}
with $H_a^0$ a proper rational transfer function which is square, monic, stable and stably invertible.
For properties of $H_b^0$ we need the following lemma, which is an adapted version of the spectral factorization theorem  \citep{Youla:61} that is also used in \cite{weerts2016identifiability2}.

\begin{lemma}[Factorization of reduced-rank spectra]
\label{lemrr}
Consider an $L$-dimensional stationary stochastic process $x$ with rational spectral density $\Phi_x$ and rank $p < L$, that satisfies the ordering property of Assumption \ref{asspi}. Then
\begin{itemize}
\item[a.] $\Phi_x$ allows a unique spectral factorization
\[ \Phi_x = F \Delta F^* \]
with $F \in \R^{L\times p}(z)$, $F = \begin{bmatrix} F_a \\ F_b \end{bmatrix}$ with $F_a$ square, monic, and $F$ stable and having a stable left inverse $F^\dagger$ that satisfies $F^\dagger F = I_p$, and $\Delta \in \R^{p\times p}$, $\Delta > 0$;
\item[b.] Based on the unique decomposition of $\Phi_x$ in (a.), there exists a unique factorization of $\Phi_x$ in the structure:
\[ \Phi_x = \breve F \breve\Delta \breve F^* \]
with $\breve F \in \R^{L \times L}(z)$ monic, stable with a stable inverse and $\breve \Delta \in \R^{L \times L}$, having the particular structure
\[
\breve F = \begin{bmatrix} F_a & 0 \\ F_b-\Gamma & I \end{bmatrix},\ \ \ \ \
 \breve \Delta = \begin{bmatrix} I \\ \Gamma \end{bmatrix} \Delta \begin{bmatrix} I \\ \Gamma \end{bmatrix}^T
\]
and $\Gamma := \lim_{z\rightarrow\infty} F_b(z)$.
\end{itemize}
\end{lemma}
{\bf Proof.} Part (a) is the standard spectral factorization theorem, see \cite{Youla:61}.
The decomposition in part (b) can be verified by direct computation.
Stability of $\check F$ follows from stability of $F$.
Stability of
\begin{equation}
	\check F^{-1} =
	\begin{bmatrix}
		F_a^{-1} & 0 \\ (F_b-\Gamma)F_a^{-1} & I
	\end{bmatrix}
\end{equation}
follows since it contains only stable components.
\hfill $\Box$

From Lemma \ref{lemrr} we know that $H^0$ is stable and has a stable left inverse $H^\dagger$, satisfying $H^\dagger H^0 = I_p$, the $p \times p$ identity matrix.
The feedthrough term of $H_b^0$ will throughout the paper be indicated with $\Gamma^0$, i.e. $\Gamma^0 := \lim_{z \rightarrow \infty} H_b^0(z).$

When we apply Lemma \ref{lemrr}b to $v(t)$ we can make a decomposition
\begin{equation} \label{eq:noisemodel}
	v(t) = \check H^0(q) \check e(t)
	= \begin{bmatrix} H_a^0(q) & 0 \\ H_b^0(q)-\Gamma^0 & I	\end{bmatrix} \begin{bmatrix} e \\ \Gamma^0 e \end{bmatrix}
\end{equation}
where $\check H^0$ satisfies the conditions in Lemma \ref{lemrr}b, and
with the $L$-dimensional white noise process $\check e$ with covariance matrix $\check \Lambda^0$ defined by:
\begin{equation} \label{eq:inno}
\check \Lambda^0 = \begin{bmatrix} I \\ \Gamma^0 \end{bmatrix} \Lambda^0 \begin{bmatrix} I \\ \Gamma^0 \end{bmatrix}^T.\end{equation}
From the definition of $\check e$ we can see that there is a particular relation between the driving white noise process in the first $p$ nodes and the last $L-p$ nodes.
This particular relation is used throughout the paper.\\
Note that with (\ref{eq:noisemodel}) there are actually two different noise model representations:
\[ v(t) = H_0(q)e(t) = \check H_0(q)\check e(t) \]
with $\check e(t)$ and $v(t)$ being $L$-dimensional, and $e(t)$ being $p$-dimensional, with $p\leq L$. In the case of full-rank noise, $p=L$ and both representations are the same. Both expressions will be utilized.

\begin{figure}[ht]
	\includegraphics[width=\columnwidth,trim={1cm 0 1cm 0},clip]{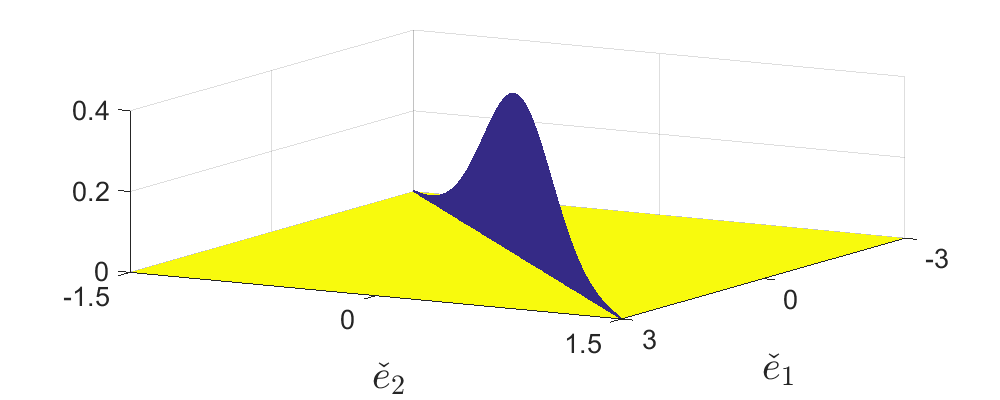}
	\caption{pdf of rank-reduced noise $\check e(t) = [e(t) \; 0.5 e(t)]^T$, with $e(t) \sim \mathcal{N}(0,\Lambda^{0})$ a 1-dimensional random variable.}
	\label{fig:pdf}
\end{figure}

The white noise process $e(t)$ is modeled as a stationary stochastic process.
The probability density function (pdf) of the rank-reduced process $\check e$ is defined by two equations \citep{rao1973}, i.e.
the pdf of $e$ and the additional constraint
\begin{equation} \label{eq:hypla}
	\begin{bmatrix}
		\Gamma^0 & -I
	\end{bmatrix}
	\check e =0.
\end{equation}
An interpretation of this characterization of $\check e$, is a $p$-dimensional pdf that lives on a plane described by (\ref{eq:hypla}).
This interpretation is illustrated in Figure \ref{fig:pdf} for an example of a 2-dimensional noise process $\check e(t)$ having rank $1$ with a Gaussian pdf.

\section{The joint-direct identification setup}
\label{Sec:predictor}
After having defined the basic network properties and representations, the next step is to formulate the identification setting. Since our goal is identify the full network dynamics, i.e. all modules $G_{ji}^0$ in the network, we are going to build a predictor model that predicts all measured node signals $w$ in the network.

\begin{definition} \label{def:predr}
The one-step-ahead predictor for node signals $w(t)$ is defined as the conditional expectation
\begin{equation}
 \hat w(t|t-1)   :=   \E \left \{ w(t) \; | \; w^{t-1},\;  r^t \right \}, \label{eq:pr1}
\end{equation}
conditioned on  $ w^{t-1} := \{w(0), w(1), \cdots, w(t-1)  \}$ and $r^t := \{r(0), r(1), \cdots, r(t)  \}.$
 \hfill $\Box$
\end{definition}

We have shown that there are multiple ways to model the noise process $v$. In order to write a unique and explicit form for the predictor filters that generate the one-step-ahead prediction, we use the squared version of the noise model (\ref{eq:noisemodel}), i.e. $v = \check H^0 \check e$. This leads to the following result.
%
\begin{proposition} \label{prop1}
For a dynamic network considered in Section \ref{Sec:NetDef},
	the one-step-ahead predictor of the node signals $w(t)$ is given by
	\begin{equation} \label{eq:pred}
		  \hat w(t|t-1) = W_w^0(q) w(t) + W_r^0(q) r(t),
	\end{equation}
	with the
    predictor filters
	\begin{align}
		&\label{eq:Ww}	W_w^0(q) = I -  (\check H^0(q))^{-1}  (I- G^0(q)), \\
		&\label{eq:Wr}	W_r^0(q) =  (\check H^0(q))^{-1}  R^0(q).
	\end{align}
\end{proposition}
\textbf{Proof:} Collected in the appendix.
\hfill $\Box$

\begin{remark}
In earlier work \citep{weerts2016identifiability} the alternative noise model, determined by $H^0(q)$ was used as a basis for formulating the predictor filters. \red{However due to intrinsic non-uniqueness of the corresponding filter expressions, the use of the square nose model $\check H^0$ is more attractive.}
Note that a subtle difference between the noise models $\check H^0$ and $H^0$ is that in $\check H^0$ the feedthrough term of $H_b^0$ has been removed and is represented now in $cov(\check e)$.
\end{remark}

In order to arrive at a network identification setup we need to specify a network model and a network model set.
\begin{definition}[network model]
\label{def1}
A network model of a network with $L$ nodes, and $K$ external excitation signals, with a noise process of rank $p \leq L$ is defined by the quadruple:
\[ M = (G,R,H,\Lambda) \]
with
\begin{itemize}
\item $G \in \R^{L \times L}(z)$, diagonal entries 0, all modules strictly proper and stable;
\item $R \in \R^{L \times K}(z)$, proper and stable;
\item $H \in \R^{L \times p}(z)$, satisfying the properties for $H^0(z)$.
\item    $\Lambda \in \R^{p\times p}$, $\Lambda > 0$;
\item the network is well-posed\footnote{The implies that all principal minors of $\lim_{z \rightarrow \infty} (I-G(z))^{-1}$ are nonzero.}  \citep{Dankers_diss}, with $(I-G)^{-1}$ proper and stable. \hfill $\Box$
\end{itemize}
\end{definition}
The data generating system is indicated by the model $\mathcal{S} = (G^0,R^0,H^0,\Lambda^0)$.

\begin{definition}[network model set] \label{def3}
A network model set for a network of $L$ nodes, $K$ external excitation signals, and a noise process of rank $p \leq L$, is defined as a set of parametrized matrix-valued functions:
\[ \M := \left\{ M(\theta) = \bigl(G(q,\theta), R(q,\theta), H(q,\theta), \Lambda(\theta)\bigr), \theta \in \Theta \right\}, \]
with all models $M(\theta)$ satisfying the properties as listed in Definition \ref{def1}.\hfill $\Box$
\end{definition}

The  data generating system $\mathcal{S}$ is represented by parameter $\theta_0$, so $\mathcal{S}=M(\theta_0)$.
In the parameterization the feedthrough of $H_b$ is modeled by $\Gamma(\theta)$ defined as $\Gamma(\theta) := \lim_{z \rightarrow \infty} H_b(z,\theta).$

Predictor (\ref{eq:pred}) will be parameterized to create the parameterized predictor
\begin{equation} \begin{split} \label{eq:pred_param}
	&\hat w(t|t-1,\theta) = w(t)  +
	\\
	&- \left ( \breve H(q,\theta) \right ) ^{-1} \left \{  (I- G(q,\theta))w(t) - R(q,\theta) r(t)  \right \},
\end{split} \end{equation}
with
\begin{equation}
	 \breve H(q,\theta) =
	 \begin{bmatrix} H_a(q,\theta) & 0 \\ H_b(q,\theta) - \Gamma(\theta) & I	 \end{bmatrix}.
\end{equation}
The prediction error is then standardly defined as
\begin{equation} \label{eq:def_prederr}
	\varepsilon(t,\theta)
	:=
	w(t)
	-
	\hat w(t|t-1,\theta),
\end{equation}
which is $L$-dimensional.

\red{For estimating the model parameters a} Weighted Least Squares (WLS) criterion is considered:
\begin{equation} \label{eq:uncon}
	\hat \theta_N^{WLS} = \arg \min_{\theta \in \Theta} \frac{1}{N} \sum_{t=1}^N \varepsilon^T(t,\theta)
	\; Q \;
	\varepsilon(t,\theta),
\end{equation}
with $Q \geq 0$.
\red{Given the multivariate character of the prediction error, the WLS criterion will allow us to show
maximimum likelihood properties, and thus asymptotic minimum variance properties of our estimated models, in Section \ref{Sec:CLS}.}
In order to analyze consistency, first we need to introduce the notion of network identifiability.
Network identifiability ensures that we are able to distinguish between different network topologies and dynamics.

\begin{definition}[Network identifiability (\cite{weerts2016identifiability2})]
\label{defif}
The network model set $\M$ is globally network identifiable at $M_0 :=M(\theta_0)$ if for all models $M(\theta_1) \in \M$,
\begin{equation} \label{equivTP}
		\left. \begin{array}{c} T_{wr}(q,\theta_1) = T_{wr}(q,\theta_0) \\ \Phi_{\bar v}(\omega,\theta_1) = \Phi_{\bar v}(\omega,\theta_0) \end{array} \right\}
		\Rightarrow
		M(\theta_1) = M(\theta_0).
\end{equation}
$\M$ is globally network identifiable if (\ref{equivTP}) holds for all $M_0 \in \M$.\hfill $\Box$
\end{definition}
Since all modules \red{in $G(q,\theta)$} are assumed to be strictly proper, condition (\ref{equivTP}) in Definition \ref{defif} is equivalently formulated as \citep{weerts2016identifiability2}:
	\begin{eqnarray} \label{equivT}
		\lefteqn{\{T(q,\theta_1) = T(q,\theta_0)\}
		\Rightarrow} \\
		& & \ \ \ \{(G(\theta_1), R(\theta_1), H(\theta_1)) = (G(\theta_0),R(\theta_0), H(\theta_0)\}, \nonumber
	\end{eqnarray}
with $T(q,\theta)$ the parametrized version of $T^0(q)$ (\ref{eq:T0}).
A sufficient condition for a model structure to be network identifiable is that every node has an independent excitation source, which can be either noise or an external excitation. This is characterized by the \red{condition} that \red{the columns of}
$\left [ H(q,\theta) \; R(q,\theta) \right ]$ \red{can be permuted to arrive at a matrix with a leading}  diagonal \citep{weerts2016identifiability2}. \red{An alternative condition that takes into account the structure of $G(\theta)$, is
presented in the following Proposition.}
%
%
\blue{\begin{proposition}[\citep{weerts2016identifiability2}] \label{thm:theo3}
Let $\M$ be a network model set that satisfies the following properties:
\begin{itemize}
\item[a.] Every parametrized entry in the model $\{ M(z,\theta), \theta\in\Theta\}$ covers the set of all proper rational transfer functions;
\item[b.] All parametrized transfer functions in the model $M(z,\theta)$ are parametrized independently (i.e. there are no common parameters).
\end{itemize}
Then $\M$ is globally network identifiable at $M(\theta_0)$ if and only if
		\begin{itemize}
			\item
			each row $i$ of the transfer function matrix\\ $\begin{bmatrix} G(\theta) & H(\theta) & R(\theta)	 \end{bmatrix}$ has at most $K+p$ parameterized entries, and
			\item
for each $i$, the transfer function from all external signals $(r_m, e_n)$ that are input to a non-parametrized module $R_{jm}(q)$ or $H_{in}(q)$, to node signals $w_k$ that are input to parameterized modules $G_{ik}(q;\theta)$, is full row rank in $\theta=\theta_0$.					\hfill $\Box$		
		\end{itemize}
\end{proposition}}
For analysis of the asymptotic properties of the parameter estimate (\ref{eq:uncon}) it is attractive to consider the asymptotic criterion
\begin{equation} \label{eq:tstar}
	\theta^\star = \arg \min_{\theta \in \Theta}  \bar V(\theta),
\end{equation}
with
\begin{equation} \label{eq:barV}
	\bar V(\theta)= \Eb \;\varepsilon^T(t,\theta)
	\; Q \;
	\varepsilon(t,\theta),
\end{equation}
and $\Eb$ defined as $ \lim_{N\rightarrow \infty}\sum_{t=1}^N \E$, according to \cite{ljung:99}.
In classical literature it has been shown that the solution of the weighted least squares criterion converges to the solution of the asymptotic criterion under some mild conditions \citep{ljung:99}. Based on this result we can formulate that,
under the condition that $w(t)$ and $r(t)$ are jointly quasi-stationary, $r(t)$ is bounded, and $e(t)$ has bounded moments of order $\geq 4$, it holds that
\begin{equation} \label{eq:conv_uncon}
	\hat \theta_N^{WLS} \rightarrow \theta^\star \text{ w.p. } 1 \text{ as } N \rightarrow \infty.
\end{equation}	
In order to show consistency, we then need to add the conditions for $\theta^\star$ to be equal to $\theta_0$. These are formulated in the next Proposition, which was presented in \cite{weerts2017ifacwc}.
\begin{proposition} \label{prop:uncon}
Let $\theta^\star$ be defined by (\ref{eq:tstar}).
Then under the conditions
\begin{enumerate}
		\item The data generating system is in the model set, i.e. $\exists \theta_0 \in \Theta$ such that $M(\theta_0) = \mathcal{S}$,
		\item external excitation $r$ is persistently exciting of sufficiently high order and uncorrelated with $e$, and
		\item $\M$ is globally network identifiable at $\mathcal{S}$,
	\end{enumerate}
	it holds that\footnote{Strictly speaking $\theta^\star$ can be a set and the equation holds for all $\theta \in \theta^\star$.}
	\begin{equation} \label{eq:2claim}   \begin{split}
		\{
			G(q,\theta^\star),H_a(q,\theta^\star),H_b(q,\theta^\star) - \Gamma(\theta^\star),R(q,\theta^\star)
		\}
		\\=
		\{
			G^0(q),H_a^0(q),H_b^0(q) - \Gamma^0,R^0(q).
		\}
	\end{split}\end{equation}
\end{proposition}
\textbf{Proof:} Collected in the appendix.
\hfill $\Box$

The matrix $\Gamma^0$ in the innovation is not estimated by the criterion (\ref{eq:uncon}),
but information on $\Gamma$ exists in the residuals.
Based on the dependencies in the innovation we split the prediction error into 2 parts:
\begin{equation}
	\varepsilon(t,\theta) = \begin{bmatrix}
		\varepsilon_a(t,\theta) \\ \varepsilon_b(t,\theta)
	\end{bmatrix},
\end{equation}
where $\varepsilon_a \in \R^p$, and $\varepsilon_b \in \R^{L-p}$.
Under zero initial conditions in the system and the predictor filters, the prediction error, when evaluated at $\theta=\theta_0$,  has the same dependencies as the innovation, i.e.
\[
	\varepsilon_a(t,\theta_0)=e(t)
	\text{,\quad and \quad}
	\varepsilon_b(t,\theta_0)=\Gamma^0 e(t),
\]
such that $\Gamma^0\varepsilon_a(t,\theta^0)=\varepsilon_b(t,\theta^0)$.
Using this knowledge an estimation of $\Gamma^0$ can be made by
\begin{equation}
	\hat \Gamma_N = \left ( \!\! \frac{1}{N} \sum_{t=1}^N \; \varepsilon_b(\hat \theta_N)\varepsilon_a^T(\hat \theta_N) \!\! \right ) \!\!\! \left ( \!\! \frac{1}{N} \sum_{t=1}^N \; \varepsilon_a(\hat \theta_N)\varepsilon_a^T(\hat \theta_N) \!\! \right )^{-1}\!.
\end{equation}
Since $\hat \theta_N$ is a consistent estimate, this estimate $\hat \Gamma_N$ will converge to
\begin{equation}
	\Gamma^\star = \big ( \E \; \varepsilon_b(\theta^\star)\varepsilon_a^T(\theta^\star) \big ) \big ( \E \; \varepsilon_a(\theta^\star)\varepsilon_a^T(\theta^\star) \big )^{-1}
\end{equation}
which is
\begin{equation}
	\Gamma^\star
	= \Gamma^0  \Lambda^0 (\Lambda^0)^{-1} = \Gamma^0.
\end{equation}

For full-rank noise the weight $Q=(\Lambda^0)^{-1}$ typically leads to minimum variance estimates, but for rank-reduced noise $\check \Lambda^0$ is not invertible.
In order to obtain minimum variance properties a new approach is needed to determine an appropriate weighting $Q$ in the identification criterion (\ref{eq:uncon}).

The identification method that has been presented here is termed as  ``joint-direct method'', as it combines elements from two classical methods for closed-loop identification \citep{ljung:99}, i.e. the joint-io method that is based on treating all measured signals jointly and starts with estimating closed-loop transfer function objects, and the direct method in which plant and noise dynamics are parametrized directly.

\section{Constrained Least Squares and Maximum likelihood}
\label{Sec:CLS}
The WLS criterion does not take into account the fact that there are dependencies in the innovation process $\check e(t)$, as represented in (\ref{eq:hypla}).
In this section we introduce an identification criterion which properly takes these dependencies into account.
It is shown that this approach leads to Maximum Likelihood estimates and to an appropriate choice of weight for the WLS criterion.
Based on the dependencies in the innovation we define
\begin{equation} \label{eq:defZ}
	Z(t,\theta) := \Gamma(\theta) \varepsilon_a(t,\theta) - \varepsilon_b(t,\theta),
\end{equation}
and introduce the Constrained Least Squares (CLS) criterion:
\begin{equation} \label{eq:crit_cons}
	\begin{split}
		\hat \theta_N^{CLS} = &\arg \min_{\theta} \frac{1}{N} \sum_{t=1}^N \varepsilon^T_a(t,\theta)
		\; Q_a \;
		\varepsilon_a(t,\theta)
		\\
		& \text{subject to } \frac{1}{N} \sum_{t=1}^N Z^T(t,\theta)Z(t,\theta)=0,
	\end{split}
\end{equation}
with $Q_a>0$.
For finite $N$, the quadratic constraint is equivalent to the constraint $Z(t,\theta)=0 \; \forall t$, which was introduced in \cite{weerts2017ifacwc}.
We have chosen for a quadratic constraint as this facilitates the convergence and consistency result in the next proposition, and because it is less computationally demanding.

While the term $\Gamma(\theta)$ was not estimated in the WLS criterion, it enters the estimation procedure now through the constraint.
Consistency of the CLS estimate can now be formulated in the next proposition of which a preliminary version was presented in \cite{weerts2017ifacwc}.

\begin{proposition} \label{prop:con}
Let $\hat \theta_N^{CLS}$ be defined by (\ref{eq:crit_cons}) and let $\theta^\ast$ be defined by
\begin{equation}
	\begin{split}
		\theta^\ast = &\arg \min_{\theta}\Eb\; \varepsilon^T_a(t,\theta)
		\; Q_a \;
		\varepsilon_a(t,\theta)
		\\
		& \text{subject to } \Eb\; Z^T(t,\theta)Z(t,\theta)=0.
	\end{split}
\end{equation}
\begin{enumerate}
\item
Under the conditions that $w(t)$ and $r(t)$ are jointly quasi-stationary, $r(t)$ is bounded, and $e(t)$ has bounded moments of order $\geq 4$,
it holds that
\begin{equation} \label{eq:conv_con}
	\hat \theta_N^{CLS} \rightarrow \theta^\ast \text{ w.p. } 1 \text{ as } N \rightarrow \infty.
\end{equation}
\item
Under conditions
\begin{enumerate}
		\item The data generating system is in the model set, i.e. $\exists \theta_0 \in \Theta$ such that $M(\theta_0) = \mathcal S$, and
		\item external excitation $r$ is persistently exciting of sufficiently high order and uncorrelated to $e$, and
		\item $\M$ is globally network identifiable at $\mathcal S$,
	\end{enumerate}
	it holds that\footnote{Strictly speaking $\theta^\ast$ can be a set and the equation holds for all $\theta \in \theta^\ast$.}
\end{enumerate}
	\begin{equation}\begin{split}
		\{
			G(q,\theta^\ast),H(q,\theta^\ast),R(q,\theta^\ast)
		\}
		=
		\{
			G^0(q),H^0(q),R^0(q)
		\}.
	\end{split}\end{equation}
\end{proposition}
\textbf{Proof:} Collected in the appendix.
\hfill $\Box$

As opposed to the consistency result for the WLS estimate in Proposition \ref{prop:uncon}, now the term $\Gamma(\theta)$, which is included in $H_b(q,\theta)$, is also estimated consistently. This estimation is taken care of by the constraint, that constrains the parameter space in order to guarantee the (static) dependency among the terms of the prediction error.
%

Motivated by the proper handling of the dependencies in the noise terms, it can be expected that the CLS estimate has a close resemblance with
the maximum likelihood estimate. This is analysed next.

\begin{theorem} \label{thm:ml}
Let $e(t)$ be normally distributed and zero mean, i.e. $ e(t) \sim \mathcal N(0,\Lambda^0)$,
and consider a parameterized model set as in Definition \ref{def3}.
Then under zero initial conditions\red{\footnote{\red{The zero initial conditions reflect values of input and output values of the predictor filters, prior to the time interval $[1,N]$, that are required to calculate the predicted node signal within the time interval.}}:}
\begin{enumerate}
\item
	The Maximum Likelihood estimate of $\theta^0$ is 	
	\begin{equation}\label{eq:ML}
	\begin{split}
		\hat \theta_N^{ML} = &\arg \max_{\theta} \log L_a(\theta)
		\\
		& \text{subject to } \frac{1}{N} \sum_{t=1}^N Z^T(t,\theta)Z(t,\theta)=0,
	\end{split}
	\end{equation}
	with
	\begin{equation} \begin{split}
	\log L_a(\theta)
	=
	\; &c - \frac{N}{2} \log \det \Lambda(\theta)
	\\ &
	- \frac{1}{2} \sum_{t=1}^N \varepsilon_a^T(t,\theta) \Lambda^{-1}(\theta) \varepsilon_a(t,\theta).
	\end{split} \end{equation}
\item
	Under the condition that $\Lambda(\theta)$ does not share parameters with $\varepsilon(t,\theta)$ the Maximum Likelihood estimate can alternatively be written as
	\begin{equation} \label{eq:ML2}
	\begin{split}
	\hat \theta_N^{ML} = &\arg \min_{\theta} \det \left ( \frac{1}{N} \sum_{t=1}^N \varepsilon_a(t,\theta)  \varepsilon_a^T(t,\theta) \right )
		\\
		& \text{subject to } 0  = \frac{1}{N} \sum_{t=1}^N Z^T(t,\theta)Z(t,\theta),
		\\
		 &\Lambda(\theta) = \frac{1}{N} \sum_{t=1}^N  \varepsilon_a(t,\theta) \varepsilon_a^{\red{T}}(t,\theta).
	\end{split}\end{equation}
\end{enumerate}
\end{theorem}
\textbf{Proof:} Collected in the appendix.
\hfill $\Box$

In (\ref{eq:ML2}) the last equation does not involve an actual constraint that limits the optimization problem, but it is merely there to specify the parameters that determine the estimated  $\Lambda$.

Note that when a model set with fixed (non-parameterized) $\Lambda$ is used, then the Maximum Likelihood estimate (\ref{eq:ML}) reduces to the Constrained Least Squares (\ref{eq:crit_cons}) estimate with $Q_a=\Lambda^{-1}$.
This implies that the CLS equipped with the appropriate weight $Q_a =  (\Lambda^0)^{-1}$ is a maximum likelihood estimator in case of Gaussian disturbances.

\begin{remark}
	If initial conditions are non-zero and not explicitly dealt with in the parametrized model, then part of the prediction error is caused by the initial conditions.
	Although this effect asymptotically goes to 0, the $\varepsilon_b$ does not have to be linearly dependent on $\varepsilon_a$, and consequently there do not exist parameters for which $Z(t,\theta)=0$ for all $t$.
	Similarly in the situation where $\M$ does not contain $\Sy$, it  is possible that there do not exist parameters for which $Z(t,\theta)=0$ for all $t$.
	When the $Z(t,\theta)$ can not be made 0 the constraint in (\ref{eq:crit_cons}) and (\ref{eq:ML}) is not feasible and the solution set of the criterion is empty.
\end{remark}

In order to deal with situations where there are non-zero initial conditions, or where the system is not in the model set, we introduce a relaxed criterion.
This relaxed criterion has a relaxed constraint, which appears as an additional penalty term, weighted by the real-valued penalty weight $\lambda>0$:
\begin{equation} \label{eq:crit_rel}
	\hat \theta_N^{rel}
	\! = \!
	\arg \min_{\theta} \! \frac{1}{N} \! \sum_{t=1}^N
	\!\!  \Big ( \!
		\varepsilon^T_a(t,\theta)
	 	Q_a
		\varepsilon_a(t,\theta)
		\! + \!
		\lambda Z^T(t,\theta) Z(t,\theta)
	\!  \Big )  \! .
\end{equation}

The above criterion is equivalent to the CLS (\ref{eq:crit_cons}) for $\lambda \rightarrow \infty$.
Another way to write the relaxed criterion is as the WLS (\ref{eq:uncon}) with parameterized weight
\begin{equation} \label{eq:Qpar}
	Q(\theta) = \begin{bmatrix}
		Q_a + \lambda \Gamma^T(\theta) \Gamma(\theta) & - \lambda \Gamma^T(\theta)
		\\
		-\lambda	\Gamma(\theta) & \lambda I
	\end{bmatrix}.
\end{equation}
Determining the optimal $\lambda$ is not the objective in this paper.
The optimal choice for $\lambda$ will depend on the contribution of initial conditions, the contribution of unmodeled dynamics and the length of the data record.

\begin{remark} \label{rem:order}
So far we have assumed that the ordering of node signals is done in a way that the first $p$ nodes are affected by a full-rank noise process, while the remaining $L-p$ nodes are affected by ``dependent'' noise. In \cite{weerts2016identifiability2} conditions have been derived \red{for appropriately ordering the noise components}. For the current situation where the modules $G_{jl}(q)$ are strictly proper, the rank $p$ and the ordering of signals can be retrieved from $T_{wr}^\infty := \lim_{z\rightarrow\infty}T_{wr}(z)$ and $\Phi_{\bar v}^\infty := \lim_{z\rightarrow\infty}\Phi_{\bar v}(z)$. This information can be estimated from data. In this paper we will assume that the requested ordering has been performed prior to the identification of the dynamics of the network.
\end{remark}

\red{
\begin{remark} \label{rem:proper}
So far we have considered the situation that all modules in $G(q,\theta)$ are strictly proper. This situation can be extended to the situation of having proper modules in $G(q,\theta)$, thus allowing direct feedthrough terms, as long as there are no algebraic loops in the network. The network identifiability results have been derived for this situation also, see \cite{weerts2016identifiability2}. The 2-node situation has been solved even for the situation of having algebraic loops \citep{weerts2016cdc}. Since the formulation of the ML result will become technically more involved for non-strictly proper modules, we have preferred to restrict to the strictly proper module situation here.
\end{remark}
}

\section{Minimum variance and the Cram\'{e}r-Rao lower bound}
\label{sect:var}
\subsection{Variance of Weighted Least Squares Estimates}
In the situation that the noise is full rank the classical reasoning on parameter variance holds \citep{ljung:99}.
For $N \rightarrow \infty$ and $\Sy \in \M$ the estimate converges under weak conditions to a normal distribution given by
\begin{equation} \label{eq:normdist}
	\underbrace{\sqrt{N} (\hat \theta_{CLS} - \theta^0)}_{:= \tilde \theta} \sim \mathcal N(0,P_\theta),
\end{equation}
with $P_\theta$ positive definite.
For full-rank noise processes, $P_\theta$ is defined by
\begin{equation} \label{eq:Pt} \begin{split}
	P_\theta =
	\left [ \Eb \psi(t) Q \psi^T(t) \right ]^{-1}
	\left [ \Eb \psi(t) Q \Lambda^0 Q \psi^T(t) \right ] \cdot	
	\\ \cdot
	\left [ \Eb \psi(t) Q \psi^T(t) \right ]^{-1},
\end{split} \end{equation}
with
\begin{equation} \label{eq:psi}
	\psi(t)
	:=
	-\frac{d}{d\theta} \varepsilon^T(t,\theta) |_{\theta=\theta_0}.
\end{equation}
For rank-reduced noise and the WLS criterion (\ref{eq:uncon}) the expression for $P_\theta$ \red{is similar to the expression above, with $\Lambda^0$ replaced by $\breve \Lambda^0$. This }
can be shown by following its derivation in \cite{Soderstrom&Stoica:89} and using $\breve \Lambda^0$ instead of $\Lambda^0$.
The variance has some lower bound $P^0_\theta$, $P_\theta \geq P^0_\theta$, which for full-rank noise is given by
\begin{equation} \label{eq:lb}
	P^0_\theta = 	\left [ \Eb \psi(t) (\Lambda^0)^{-1} \psi^T(t) \right ]^{-1}.
\end{equation}
However for the rank-reduced case, ${\Lambda^0}$ would have to be replaced by $\breve \Lambda^0$, which, however, is singular, and so its inverse does not exist.
Therefore a lower bound like (\ref{eq:lb}) is not valid in this case.
Here is a simple example to illustrate this point.

\begin{example} \label{exa_rr}
\begin{figure}[htb]
	\includegraphics[width=\columnwidth]{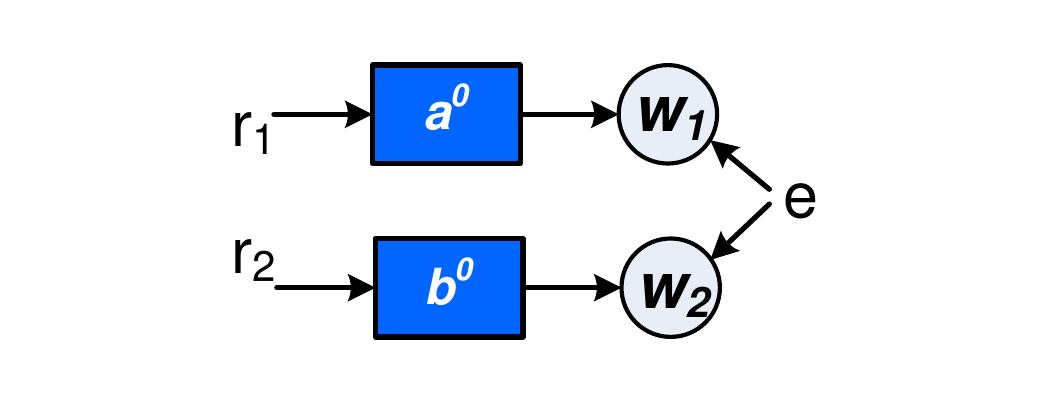}
	\caption{System with 2 nodes, no dynamics and 1 noise disturbance. It is excited by the quasi-stationary excitation signals $r_1, r_2$ and the stochastic process $e$ which are all mutually uncorrelated and have unit variance.}
\label{fig:simple}
\end{figure}
	Consider the system in Figure \ref{fig:simple}, where
	2 parameters are to be estimated, $\theta_a$ and $\theta_b$.
	The system is governed by
	\[
		\begin{bmatrix}
			w_1(t) \\ w_2(t)
		\end{bmatrix}
		=
		\begin{bmatrix}
			a^0 & 0 \\ 0 & b^0
		\end{bmatrix}
		\begin{bmatrix}
			r_1(t) \\ r_2(t)
		\end{bmatrix}
		+\underbrace{
		\begin{bmatrix}
			e(t) \\ e(t)
		\end{bmatrix}}_{\check e(t)}.
	\]
	The disturbance process $\check e$ has covariance matrix $ \check \Lambda^0 = \left [ \begin{smallmatrix}	1&1\\1&1 \end{smallmatrix} \right ]$,
	which is singular.	
	When the WLS (\ref{eq:uncon}) is used with a weight $Q$ defined by (\ref{eq:Qpar}) and $\Gamma(\theta)$ set to 1,
	\begin{equation}
		Q = \begin{bmatrix}
		1 + \lambda & -\lambda \\ -\lambda &  \lambda
		\end{bmatrix}
	\end{equation}
	with $\lambda > 0$ and
	prediction errors
	\begin{equation}
		\varepsilon_1 = w_1 - \theta_a r_1, \quad \varepsilon_2 = w_2 - \theta_b r_2,
	\end{equation}	
	then we get a consistent estimate.
	\red{The identificaiton criterion to be minimized becomes}
	\begin{equation}
		\red{\frac{1}{N}\sum_{t=1}^{N}} \left\{\frac{1}{\lambda}\varepsilon_a^2(\theta_a)+ \left (	\begin{bmatrix}
			1 & -1
		\end{bmatrix}
		\begin{bmatrix}
			\varepsilon_a(\theta_a) \\ \varepsilon_b(\theta_b)
		\end{bmatrix} \right )^2 \right\}.
	\end{equation}	
	\red{In the limit as $\lambda\rightarrow\infty$, $Q$ becomes singular, and the expression for the identification criterion reduces to}
	\begin{equation}
		\red{\frac{1}{N}\sum_{t=1}^{N}} \left\{(a^0- \theta_a)r_1 + e - (b^0- \theta_b)r_2 - e\right\}^2.
	\end{equation}
	\red{In this expression} the disturbance $e$ drops out, and a variance-free estimate of $a^0$ and $b^0$ is obtained.

This phenomenon should also be observed in the 	
variance expression (\ref{eq:Pt}). \red{If we denote $\theta = [\theta_a\ \theta_b]^T$ then}
	\begin{equation} \psi(t) = \begin{bmatrix}
	r_1(t) & 0 \\0 & r_2(t)
	\end{bmatrix},		\end{equation}
	and we obtain
	\begin{equation}
		\Eb \psi Q \psi^T = \Eb \begin{bmatrix}
			r_1^2 (1+\lambda) & -r_1r_2\lambda \\ -r_1r_2\lambda & r_2^2 \lambda
		\end{bmatrix}
		= \begin{bmatrix}
			 (1+\lambda) & 0 \\ 0 & \lambda
		\end{bmatrix}
	\end{equation}
	and
	\begin{equation} \label{eq:var1}
		\Eb \psi Q \check \Lambda^0  Q \psi^T =  \Eb \begin{bmatrix}
			r_1^2  & 0 \\ 0 & 0
		\end{bmatrix} = \begin{bmatrix}
		1&0\\0&0
		\end{bmatrix}.
	\end{equation}
	We can compute $P_\theta$ of (\ref{eq:Pt}) as
	\begin{equation}
		P_\theta =
		\begin{bmatrix}
			 (1+\lambda) & 0 \\ 0 & \lambda
		\end{bmatrix}^{-1}
		\begin{bmatrix}
		1&0\\0&0
		\end{bmatrix}
		\begin{bmatrix}
			 (1+\lambda) & 0 \\ 0 & \lambda
		\end{bmatrix}^{-1}
		=
		\begin{bmatrix}
			\frac{1}{(1+\lambda)^2} & 0 \\ 0 & 0
		\end{bmatrix}
		.
	\end{equation}
	Here we can see that as $\lambda \rightarrow \infty$ the covariance goes to 0.
	 \red{This phenomenon of variance-free estimation has also been observed in \cite{everitt2015effect}.}
\end{example}

It could be tempting to use an expression like (\ref{eq:lb}) for the lower bound on the variance, with the inverse covariance $(\Lambda^0)^{-1}$ replaced by a pseudo-inverse of $\Lambda^0$. In this example
$(\check \Lambda^0)^\dagger = \frac{1}{4} \left [ \begin{smallmatrix} 1&1\\1&1 \end{smallmatrix} \right ]$ and substituting this into (\ref{eq:lb}) in stead of $(\Lambda^0)^{-1}$, delivers
	\begin{equation}
		\left [ \Eb \psi(t) (\check \Lambda^0)^\dagger \psi^T(t) \right ]^{-1} \neq 0,
	\end{equation}
	which can not be the expression for the minimum variance.

	Note that this example is fully symmetric in nodes $w_1$ and $w_2$, or equivalently in systems $a^0$ and $b^0$.
	Nevertheless one of the parameters \red{$\theta_b$} is estimated variance-free, while \red{$\theta_a$} is not.
	This is the result of the particular choice of weighting function, that according to (\ref{eq:Qpar}) reflects the choice of $w_1$ as the full-rank noise node.
	Choosing the alternative weight
	$Q = \left [ \begin{smallmatrix} \lambda & -\lambda \\ -\lambda & 1+\lambda \end{smallmatrix} \right ]$
	would resemble the situation of choosing $w_2$ as the full rank noise node,
	For both the weights, when we let $\lambda\rightarrow\infty$ the variance-free maximum likelihood estimate is obtained, which is again symmetric in $\theta_a$ and $\theta_b$.

In this example it is possible to choose a weight beyond the structure of (\ref{eq:Qpar}), e.g.
$Q = \left [ \begin{smallmatrix} 1 & -1 \\ -1 & 1 \end{smallmatrix} \right ]$, in which case we arrive at a variance-free estimate, since $Q\Lambda^0 Q=0$.
For this choice of $Q$ we are essentially only modeling the 'constraint', and we dropped the 'original cost function' $\varepsilon_a^2$.
Such a weight $Q$ is useful when all parameters in the model can be estimated using just the constraint.

\subsection{Variance of Constrained Least Squares Estimates}

In order to find a closed-form expression for the minimum variance in the case of the CLS estimate, we have to follow a different route, and address the full impact of the constraint, that typically reduces the effective parameter space in the criterion.

For the CLS situation we will write the asymptotic identification criterion as follows:
\begin{equation} \begin{split}
	\theta^\ast = &\arg \min_{\theta}\Eb\; \varepsilon^T_a(t,\theta)
		\; Q_a \;
		\varepsilon_a(t,\theta)
		\\
		&\text{subject to: } f(\theta) = 0,
\end{split} \end{equation}
with $f(\theta)$ defined by $f(\theta) = \Eb \; Z^T(t,\theta)Z(t,\theta)$.
We are making an analysis of the variance around the true system parameters, so we assume that $\theta^\ast=\theta^0$.

\red{
In $\theta = \theta^\ast$ we can represent the constraint using a first order Taylor series approximation of $Z$ in $\theta^\ast$,
i.e.
\begin{equation}
 	Z(t,\theta) \approx
 	Z(t,\theta^\ast) +
 	\frac{\partial Z(t,\theta)}{\partial \theta} \Big |_{\theta=\theta^\ast}   (\theta - \theta^\ast)
 	=  A(t) (\theta - \theta^\ast),
\end{equation}
where
\begin{align}
	A(t) &:= \frac{\partial Z(t,\theta)}{\partial \theta} \Big |_{\theta=\theta^\ast}
\end{align}
with $A \in \R^{(L-p) \times n_\theta}$.
The approximated constraint is then
\begin{equation}
 	 \Eb \;  (\theta - \theta^\ast)^T A^T(t)  A(t) (\theta - \theta^\ast) =0,
\end{equation}
where $\Eb \; A^T(t)  A(t)$ is of dimension $n_\theta \times n_\theta$ and is of rank $(n_\theta-n_\rho)$ the number of degrees of freedom that the constraint removes.
We can define a matrix $\Pi$ of dimension $(n_\theta-n_\rho) \times n_\theta$ of full row rank such that the expectation is
\begin{equation}
 	\Eb \; A^T(t)  A(t) = \Pi^T \Pi.
\end{equation}
Then in the neighborhood of the estimate $\theta^\ast$ the constraint is approximated by a quadratic constraint
\begin{equation} \label{eq:crit_appr} \begin{split}
	\theta^\ast = &\arg \min_{\theta}\Eb\; \varepsilon^T_a(t,\theta)
		\; Q_a \;
		\varepsilon_a(t,\theta)
		\\
		&\text{subject to: }    (\theta - \theta^\ast)^T \Pi^T \Pi (\theta - \theta^\ast) =0.
\end{split} \end{equation}
}

In order to appropriately take the constraint into account in the variance analysis, a re-parameterization will be considered using a parameter $\rho$ with dim$(\rho) = n_\rho < $ dim$(\theta)$.
The two parameters will be related through a mapping induced by the constraint, such that the new parameterization trivially satisfies the constraint.
\begin{lemma} \label{lem:rewrite_constrain}
	The constrained parameter space determined by \red{$  (\theta - \theta^\ast)^T \Pi^T \Pi (\theta - \theta^\ast) =0$, with $\Pi$ defined as above}, is equivalently described by
	\begin{equation} \label{eq:mapping}
		\theta = S \rho + C, \quad \text{with } \rho \in \R^{n_\rho},
	\end{equation}
	where $S \in \R^{n_\theta \times n_\rho}$ satisfies \red{$\Pi S=0$} and is full column rank, i.e. $S$ characterizes the right nullspace of \red{$\Pi$,
	and $C = \Pi^\dagger \Pi \theta^\ast$, where $\Pi^\dagger$ satisfies $\Pi \Pi^\dagger =I$.}
\end{lemma}
\textbf{Proof:} Collected in the appendix.

The unconstrained parameter $\rho$ can now
be used to rewrite the criterion (\ref{eq:crit_appr}) into a form that trivially satisfies the constraint.
The resulting criterion is then essentially an unconstrained criterion operating on a lower dimensional parameter $\rho$.

\begin{proposition}
	\label{prop:crit_rho}
	The optimization problem (\ref{eq:crit_appr}) can equivalently be written as
	\begin{equation} \label{eq:rho}
		\theta^\ast = S \rho^\ast +C,
	\end{equation}		
	with
	\begin{equation} \label{eq:crit_rho}
		\rho^\ast = \arg \min_{\rho} \bar{\E} \;
		\varepsilon_a(t,S\rho+C)
		\; Q_a \;
		\varepsilon_a(t,S\rho+C).
	\end{equation}
\end{proposition}
\textbf{Proof:} Collected in the appendix.
\hfill $\Box$

Since (\ref{eq:crit_rho}) is an unconstrained identification criterion, we know that the asymptotic variance of  the estimate $\hat \rho_N$ that corresponds to the asymptotic estimate $\rho^\ast$ is given by
\begin{equation} \label{eq:Pr} \begin{split}
	P_\rho =
	\left [ \Eb \psi_\rho(t) Q_a \psi_\rho^T(t) \right ]^{-1}
	\left [ \Eb \psi_\rho(t) Q_a \Lambda^0 Q_a \psi_\rho^T(t) \right ] \cdot	
	\\ \cdot
	\left [ \Eb \psi_\rho(t) Q_a \psi_\rho^T(t) \right ]^{-1},
\end{split} \end{equation}
with
\begin{equation} \label{eq:psirho}
	\psi_\rho(t)
	=
	-\frac{d}{d\rho} \varepsilon^T_a(t,S \rho +C) |_{\rho=\rho^\ast}.
\end{equation}
Combining this expression with (\ref{eq:mapping}) now provides an expression for $P_\theta$, as formulated next.
\begin{proposition}
	\label{prop:relmap}
	The covariance matrices $	P_\rho $ and $ P_\theta$ satisfy the following relation
	\begin{equation}
		P_\theta = S P_\rho S^T.
	\end{equation}
\end{proposition}
\textbf{Proof:} Collected in the appendix.
\hfill $\Box$

It is well known that the lower bound of $P_\rho$ is achieved when $Q_a=(\Lambda^0)^{-1}$, such that
\begin{equation}
	P_\rho \geq P_\rho^0 =
	\left [ \Eb \psi_\rho(t) (\Lambda^0)^{-1} \psi_\rho^T(t) \right ]^{-1}.
\end{equation}
Then by Proposition \ref{prop:relmap} the lower bound of $P_\theta$ is achieved by
\begin{equation} \label{eq:lbt}
	P_\theta \geq P_\theta^0 = S P_\rho^0 S^T.
\end{equation}
So the lower bound of $P_\theta$ is achieved for $Q_a=(\Lambda^0)^{-1}$.
Matrix $S$ characterizes the right-nullspace of $\Pi$, so it is not a unique matrix, but all possible $S$ matrices lead to the same $P_\theta$ and lower bound.
As an illustration of the results, an Example is shown for the CLS estimate.

\begin{example}
\begin{figure}[htb]
	\includegraphics[width=\columnwidth]{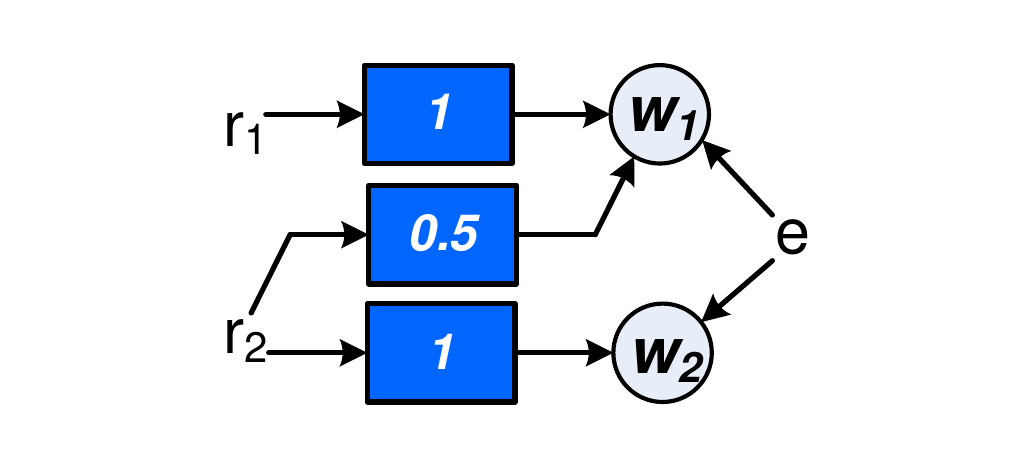}
	\caption{System with 2 nodes, no dynamics and 1 noise disturbance. It is excited by the quasi-stationary excitation signals $r_1, r_2$ and the stochastic process $e$ which are all mutually uncorrelated and have unit variance.}
	\label{fig:simple3}
\end{figure}
	In this example, depicted in Figure \ref{fig:simple3}, the system is given by
	\begin{equation}
		w_1(t) = r_1(t) + 0.5 r_2(t) +e(t) , \quad w_2(t) =  r_2(t) +e(t).
	\end{equation}
	The noise is rank reduced, and has covariance matrix $ \check \Lambda^0 = \left [ \begin{smallmatrix}	1&1\\1&1 \end{smallmatrix} \right ]$,
	which is singular.	
	When the CLS (\ref{eq:crit_cons}) is used with knowledge of $\Gamma^0=1$ and
	prediction errors
	\begin{equation}
		\varepsilon_1 = w_1 - \theta_{a1} r_1- \theta_{a2} r_2, \quad \varepsilon_2 = w_2 - \theta_b r_2,
	\end{equation}	
	where $\varepsilon_1=\varepsilon_a$ and $\varepsilon_2=\varepsilon_b$, then we get a consistent estimate.
	The constraint here consists of \red{$\Eb \; Z^2(t,\theta) =0$ with
	\begin{equation}
	 	Z(t,\theta)  = \varepsilon_1(t,\theta) - \varepsilon_2(t,\theta).
	\end{equation}}

	\red{
	With $Z(t,\theta)$ being linear in $\theta$ it follows easily that
	\begin{equation}
	 		A(t) = \frac{\partial Z(t,\theta)}{\partial \theta} \Big |_{\theta=\theta^\ast}
	 		= \begin{bmatrix}
			- r_1 (t) & - r_2(t) &  r_2(t)
		\end{bmatrix}.
	\end{equation}
	When evaluating the expectation in the constraint we have
	\begin{equation}
	 	\Eb A^T(t) A(t) = \Eb
	 	\begin{bmatrix}
	 		r_1^2(t) &0&0\\
	 		0& r_2^2(t) & -r_2^2(t) \\
	 		0& -r_2^2(t) &r_2^2(t)
	 	\end{bmatrix}
	 	=
	 	\begin{bmatrix}
	 		1&0&0\\0&1&-1\\0&-1&1
	 	\end{bmatrix},
	\end{equation}
	such that $\Pi = \begin{bmatrix}
		1&0&0\\0&1&-1
	\end{bmatrix}$.}

\red{
Vectors $S$ and $C$ can now be determined based on $\Pi S=0$ and $C=\Pi^\dagger \Pi \theta^\ast$,}
leading to:
	\begin{align*}
		S = \begin{bmatrix}
		0\\1\\ 1
		\end{bmatrix}, \quad
		C = \begin{bmatrix}
			1 \\ -0.5 \\ 0
		\end{bmatrix}.
	\end{align*}
	With this choice of $S$, we can determine $\psi_\rho$ using (\ref{eq:psirho}) as
	\begin{equation}
		\psi_\rho = - \frac{d}{d\rho}
		\left (
		w_1 -\begin{bmatrix}
			r_1 & r_2 & 0
		\end{bmatrix}
		(S\rho +C) \right )
		=
		r_2
	\end{equation}
	Then	$P_\rho$ of (\ref{eq:Pr}) is given by:
	\begin{align*}
		P_\rho = (\Eb \; r_2^2)^{-1} (\Eb \; r_2^2) (\Eb \; r_2^2)^{-1} =1,
	\end{align*}
	where $\Lambda^0=Q_a=1$.
	Then with Proposition \ref{prop:relmap} the covariance of $\theta$ is determined as
	\begin{align*}
		P_\theta = S P_\rho S^T = \begin{bmatrix}
		0&0&0 \\0&1&1\\0&1&1
		\end{bmatrix}.
	\end{align*}
Since we used the optimal weighting $Q_a=(\Lambda^0)^{-1}$ this is also the lower bound on the variance in the given situation.
Note that in the considered situation the first parameter $\theta_{a_1}$ is estimated variance-free.
\end{example}

\begin{remark}
The lower bound on the variance can be at 0, in particular situations.
When the matrix $\Eb  A^T(t) A(t)$ in the constraint is square and full rank, then the constraint uniquely determines all parameters, and {\it all} parameters are determined variance-free.
\end{remark}

Using a different reasoning than presented above, in \cite{stoica1998cramer} the Cram\'{e}-Rao lower bound on the variance under parametric constraints has been derived for Gaussian distributed noise.
That result can be linked to the lower bound that we just obtained.
In \cite{stoica1998cramer} it is stated that first the Fisher information matrix $J$ of the unconstrained part of the criterion (\ref{eq:crit_cons}) is obtained, which is
\begin{equation}
	J = \bar{\E} \; \psi_a(t) \Lambda_0^{-1} \psi_a^T(t),
\end{equation}
with  $\psi_a(t) = \psi(t) \left [ \begin{smallmatrix} I \\ 0 \end{smallmatrix}\right ]$.
This unconstrained part of the criterion does not contain all parameters, meaning that $\psi_a$ contains rows that are 0, and $J$ is singular.
The lower bound on the variance can not be given by $J^{-1}$ since it does not exist.
In \cite{stoica1998cramer} it has been proven that the lower bound is given by
\begin{equation} \label{eq:CCRB}
	P_\theta^0 =
	S \left (
	S^T \bar{\E} \; \psi_a(t) \Lambda_0^{-1} \psi_a^T(t) S
	\right )^{-1} S^T,
\end{equation}
with $S$ as defined before.
The above expression is equal to the lower bound in (\ref{eq:lbt}) that we obtained using a different reasoning, since by the chain rule for differentiation we have that
\begin{equation}
	\psi_\rho(t) = S^T \psi_a(t)
\end{equation}
which can be substituted in (\ref{eq:CCRB}) to arrive at (\ref{eq:lbt}).

\section{Simulation example}
\label{Sec:simulation}
In this simulation example a 3 node network will be identified from data using the WLS and CLS criteria.
We use the network in Figure \ref{fig:example_sys} with $r_1=0$ and $v$ a 2-dimensional white noise process with $\Lambda^0=I$, such that
\begin{align*}
		G^0 = \begin{bmatrix}
			0 & G_{12}^0 & G_{13}^0 \\ 0 & 0 & G_{23}^0 \\ G_{31}^0 & 0 & 0
		\end{bmatrix},
		\quad
		H^0 = \begin{bmatrix}
			1 & 0 \\ 0 & 1 \\ 0 & 1
		\end{bmatrix}.
\end{align*}
The dynamic modules are finite impulse responses with the following coefficients
\begin{equation*}
	\begin{bmatrix}
		G^0_{12}(q) \\ G^0_{13}(q) \\ G^0_{23}(q) \\ G^0_{31}(q)
	\end{bmatrix}
	=
	\begin{bmatrix}
		0.33 & -0.2 & 0.13 & -0.08 & 0.05
		\\
		0.2 & -0.45 & -0.73 & -0.54 & -0.25
		\\
		-0.15 & 0.12 & -0.9 & 0.6 & 0.3
		\\
		-0.5 & 0.06 & -0.1 & 0.03 & 0
	\end{bmatrix}
	\begin{bmatrix}
		q^{-1} \\ q^{-2} \\ q^{-3} \\ q^{-4} \\ q^{-5}
	\end{bmatrix} \!.
\end{equation*}
In total 100 Monte-Carlo simulations are performed on the above network with $N=1000$ samples taken for each data set.

A model structure is used with $G(q,\theta)$ having the same structure as $G^0$, $H(q,\theta)= \left [ \begin{smallmatrix} I \\ \Gamma(\theta_\Gamma) \end{smallmatrix} \right ]$, and with $\Lambda=I$.
Parameters are collected in the vector
\begin{equation}
	\theta^T = \begin{bmatrix}
		\theta_{12}^T & \theta_{13}^T & \theta_{23}^T & \theta_{31}^T & \theta_\Gamma^T
	\end{bmatrix} \ \in \R^{22},
\end{equation}
where $\theta_{ij}$ correspond to module $G_{ij}(\theta_{ij})$.
The prediction error can be denoted by
\begin{equation}
	\begin{bmatrix}
		\varepsilon_1(t,\theta) \\ \varepsilon_2(t,\theta) \\ \varepsilon_3(t,\theta)
	\end{bmatrix}
	=
	\begin{bmatrix}
		w_1(t) \\ w_2(t) \\ w_3(t)
	\end{bmatrix}
	-
	\begin{bmatrix}
		0 & \phi_2(t) & \phi_3(t) & 0 \\ 0&0&\phi_3(t)&0 \\ \phi_1(t)&0&0&0
	\end{bmatrix}
	\theta,
\end{equation}
with appropriately chosen regressors $\phi_i(t)$.

\begin{figure}[htb]
	\includegraphics[width=\columnwidth,trim={1cm 1cm 1cm 1cm},clip]{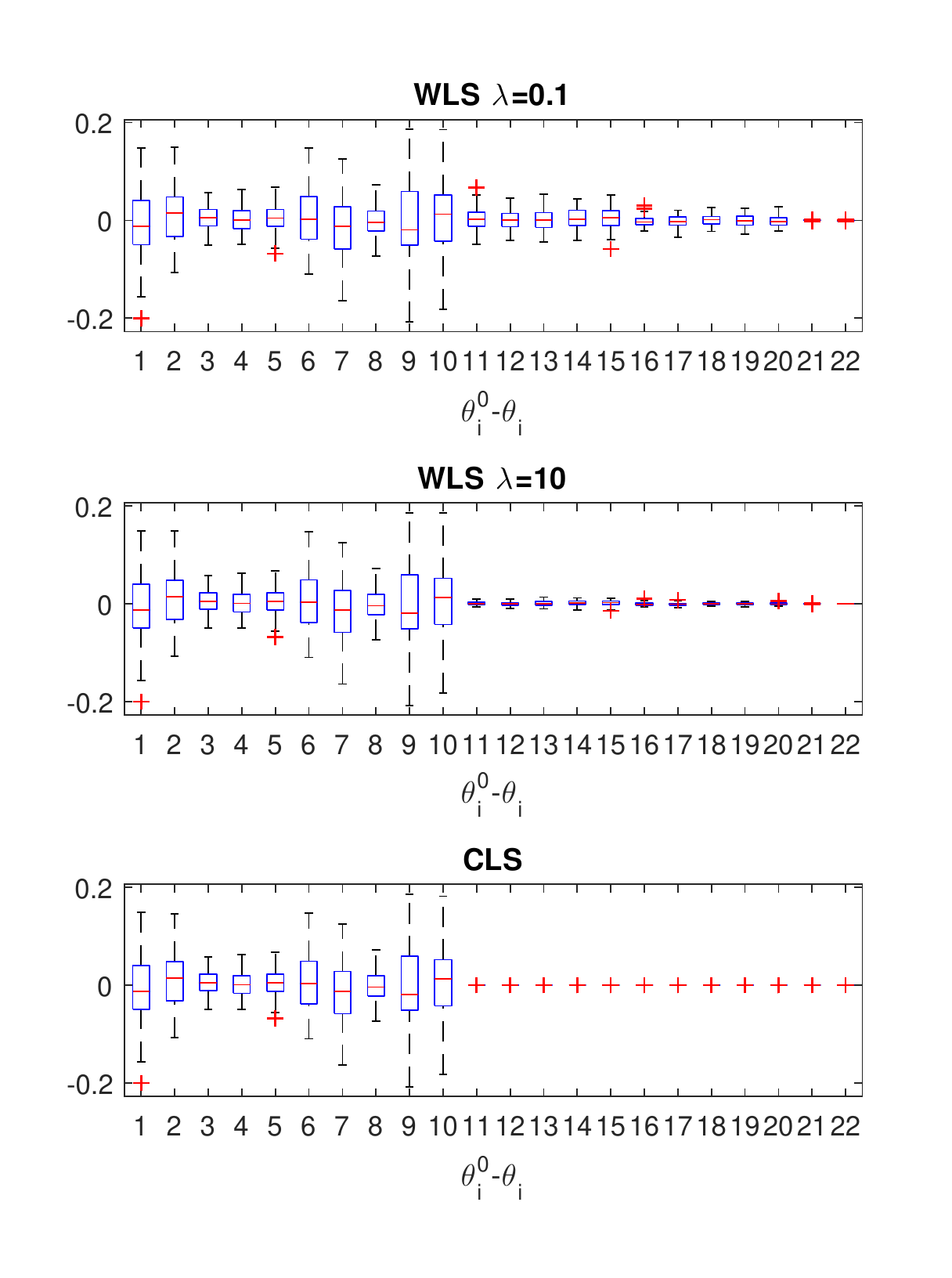}
	\caption{Boxplot of parameter estimation errors for the $22$ different parameters over 100 Monte-Carlo runs. The top and middle figures are the WLS estimates (\ref{eq:uncon}) with weight (\ref{eq:Qpar}) and $\lambda=0.1$ and $\lambda=10$ respectively,
	the bottom figure is the CLS estimate (\ref{eq:crit_cons}).}
	\label{fig:tvr}
\end{figure}

The WLS is applied as the relaxed CLS with weight (\ref{eq:Qpar}) parameterized with $\Gamma(\theta)$.
Two different choices for $\lambda$ are used to illustrate the effect of increasing values of $\lambda$.
Results of the WLS estimates, and of the CLS estimates, are plotted in Figure \ref{fig:tvr}.

It can be observed that the parameters of modules $G_{12}$ and $G_{13}$ do not change with different criteria, as the noise on node 1 is independent from noise on nodes 2 and 3.
The parameters of $G_{23}$ and $G_{31}$ are estimated with smaller variance when $\lambda$ increases, since the estimate gets closer to the ML estimate.
The parameters of $\Gamma$ (indexed by numbers 21 and 22) are estimated with very small variance, even for small $\lambda$.

For this estimation the lower bound on the variance can be computed.
\red{The constraint is based on
\begin{equation}
Z(t,\theta) = \Gamma_1(\theta) \varepsilon_1(t,\theta)	+ \Gamma_2(\theta)\varepsilon_2(t,\theta)- \varepsilon_3(t,\theta),
\end{equation}
and its derivative with respect to the parameters:
\begin{equation}\begin{split}
	&A(t) := \frac{d}{d\theta} Z(t,\theta) \Big |_{\theta=\theta^0}=
	\\
	&-\begin{bmatrix}
		\Gamma_1^0 \phi_2 & \Gamma_1^0 \phi_3 & \Gamma_2^0 \phi_3 & -  \phi_1 & \phi_2 \theta_{12}^0+\phi_3 \theta_{13}^0 & \phi_3 \theta_{23}^0
	\end{bmatrix},
\end{split}\end{equation}}
where $\Gamma^0 = \left [ \begin{smallmatrix} \Gamma_1^0 & \Gamma_2^0 \end{smallmatrix} \right ]$.

\red{
For constructing the constraint based on finite time data, the expression $\Eb A^T(t)A(t)$ is replaced by its sampled version, $\frac{1}{N} \bar A^T \bar A$ with $\bar A^T := \begin{bmatrix} A(1)^T & \cdots A(N)^T  \end{bmatrix}$. The full rank matrix $\Pi$ can then be constructed by taking an $(n_\theta - n_\rho) \times n_\theta$ full row rank sub-matrix of $\bar A$, according to
\begin{equation}
	\Pi = \begin{bmatrix}
		A(1)
		\\
		\vdots
		\\
		A(12)
	\end{bmatrix}.
\end{equation}}
Because of the fact that $\Gamma_1^0=0$, the $\red{\Pi}$ matrix is structured such that the left most 10 columns are 0.
The other 12 columns constitute a $12 \times 12$ matrix of full rank.
Matrix $S$ is then defined by the right-nullspace of \red{$\Pi$}, and $S$ has the particular structure that the first 10 rows are non-zero and form a $10 \times 10$ matrix of full rank, and the other 12 rows are 0 such that $\red{\Pi} S=0$.
When we consider $P_\theta^0=S P_\rho^0 S^T$ and the structure of $S$, it is immediately observed that the lower bound on the variance of parameters 11 to 22 is 0.

\red{The Example shows a similar phenomenon as the static Example \ref{exa_rr} in Figure \ref{fig:simple}, i.e. two modules that map into node variables that are subject to the same disturbance, and as a result of this are estimated variance-free.}

\section{Conclusions}
\label{Sec:concl}
For dynamic networks with rank-reduced noise, an appropriately parameterized model combined with a weigthed least squares criterion leads to consistent estimates under standard conditions. However for arriving at minimum variance and maximum likelihood results (under Gaussian disturbances), the required identification criterion becomes a weighted quadratic criterion subject to a constraint. A classical variance expression can be derived for the weighted least squares estimator, but for the criterion with constraint the expressions need to be modified to appropriately deal with the constraint. For this latter situation explicit expressions for the variance have been derived, as well as expressions for the lower bound of this variance, reaching the Cram\'{e}r-Rao lower bound for normally distributed noise. It has been observed and explained that parameters can be estimated variance-free. The analytical results have been illustrated with simulation examples.


\appendix
\section{Proof of Proposition \ref{prop1}}
First one predictor expression is derived using the square and monic noise model $\check H^0$, then it is shown that this is unique.
We write the network equation (\ref{eq.dgsMatrix}) as
\[ w = G^0w + R^0 r + (\check H^0-I) \check e + \check e. \]
Then we substitute using $He=\check H \check e$ and (\ref{eq.dgsMatrix}) the expression
\[ \check e = (\check H^0)^{-1}[(I-G^0)w - R^0 r] \]
into the expression $(\check H^0-I)\check e$, leading to
\begin{equation} \label{eq:preder1}
	w = [I-(\check H^0)^{-1}(I-G^0)]w + (\check H^0)^{-1}R^0 r + \check e.
\end{equation}
Since we assume that $G^0$ is strictly proper, $[I-(\check H^0)^{-1}(I-G^0)]$ is strictly proper, and evaluating the conditional expectation (\ref{eq:pr1}) leads to (\ref{eq:pred}).


\section{Proof of Proposition \ref{prop:uncon}}
First it will be shown that $\theta_0$ is a minimum of the criterion, i.e. $\theta^0 \in \theta^\star$, after which it will be shown that $M(\theta_0)$ is the only minimum, i.e. $M(\theta_0) = M(\theta) \; \forall \; \theta \in \theta^\star$.

When combining (\ref{eq:def_prederr}), (\ref{eq:pred_param}) and (\ref{eq.dgsMatrix}) it can be shown that the prediction error can be rewritten in terms of $e$ and $r$
\begin{equation} \label{eq:eer}
	\begin{bmatrix} \varepsilon_a(\theta) \\ \varepsilon_b(\theta)	\end{bmatrix} =
	F_e(q,\theta) e
	+ \begin{bmatrix} I \\ \Gamma^0	\end{bmatrix} e +
	F_r(q,\theta) r,
\end{equation}
with
\begin{align*}
	F_e(\theta) &:=
		\check H^{-1}(\theta)
		(I-G(\theta))
		(I-G^0)^{-1}H^0
		- \left [ \begin{smallmatrix} I \\ \Gamma^0	\end{smallmatrix}\right ] ,
	\\
	F_r(\theta) &:=
	\check H^{-1}(\theta)
	\Big ( (I-G(\theta))(I-G^0)^{-1} R^0 - R(\theta) \Big ),
\end{align*}
where $F_e$ is strictly proper since the innovation $\left [ \begin{smallmatrix} I \\ \Gamma^0	\end{smallmatrix}\right ] e$ has been written as a separate term.

The first term has a strictly proper filter,
the innovation (second) term does not have delay,
and since $e$ is a white noise,
the first 2 terms are uncorrelated with each other.
By condition \emph{2} the $r$ term is uncorrelated with the $e$ terms.
In the quadratic function $\bar V(\theta)$ defined by (\ref{eq:barV})
 any cross-term between the 3 terms is 0 due to uncorrelatedness, therefore each of the terms can be minimized individually.
Due to condition \emph{1} the first and third terms are minimized by $\theta_0$ and become 0.
The second term does not contain parameters, so it is trivially minimized.
Then we can conclude that $\theta^0 \in \theta^\star$.

Now it will be shown that any parameter $\theta_1$ which reaches the minimum of the cost function must result in $M(\theta_0)=M(\theta_1)$.
It can be shown  (\citep{ljung:99} proof of Theorem 8.3)   that
\begin{equation}  \begin{split}
	0	&= \bar V(\theta_0) - \bar V(\theta_1)
	\\ &= \Eb
	(\varepsilon(t,\theta_0) - \varepsilon(t,\theta_1))^T Q (\varepsilon(t,\theta_0) - \varepsilon(t,\theta_1)).
\end{split} \end{equation}
Since $Q>0$ we must have $\varepsilon(t,\theta_0) = \varepsilon(t,\theta_1)$, up to a possible transient term due to initial conditions, which decays to zero and therefore can be neglected in our asymptotic criterion.
By condition \emph{2}, $\left [ \begin{smallmatrix} e(t) \\ r(t)	\end{smallmatrix} \right ]$ is a full rank process, such that
\begin{equation}
	F_e(q,\theta_0) + \begin{bmatrix} I \\ \Gamma^0	\end{bmatrix}
	=
	F_e(q,\theta_1) + \begin{bmatrix} I \\ \Gamma^0	\end{bmatrix}
\end{equation}
and
\begin{equation}
	F_r(q,\theta_0) = F_r(q,\theta_1).
\end{equation}
Since $F_e(q,\theta_0)=0$ and $F_r(q,\theta_0)=0$ we can write
\begin{equation} \label{eq:prop2eq}
	\begin{bmatrix}
		I & 0 \\ \Gamma^0 & 0
	\end{bmatrix}
	=
	\begin{bmatrix}
		I & 0 \\ \Gamma^0 & 0
	\end{bmatrix}
	+
	\begin{bmatrix}
		F_e(q,\theta_1) & F_r(q,\theta_1)
	\end{bmatrix}.
\end{equation}
When we use the expressions for $F_e$ and $F_r$, then pre-multiply both sides of (\ref{eq:prop2eq})  with $(I-G(q,\theta_1))^{-1} \check H(q,\theta_1)$, and finally add $\begin{bmatrix} 0 &  (I-G(q,\theta_1))^{-1} R(q,\theta_1) \end{bmatrix}$ to both sides, then
\begin{equation} \label{eq:2TT} \begin{split}
	&(I-G(\theta_0))^{-1}
	\begin{bmatrix} H(\theta_0) & R(\theta_0) \end{bmatrix}		
	 = \cdot \\ \cdot
	 &\underbrace{
	(I-G(\theta_1))^{-1} \begin{bmatrix}
		H_a(\theta_1) & R_a(\theta_1) \\ H_b(\theta_1) - \Gamma(\theta_1) +\Gamma^0 & R_b(\theta_1)
	\end{bmatrix}
	}_{:=T'(\theta_1)}
\end{split} \end{equation}	
is obtained, where $R_a$ and $R_b$ are defined by $R(q,\theta) = \left [ \begin{smallmatrix} R_a(q,\theta) \\ R_b(q,\theta) \end{smallmatrix} \right ]$.
Note that $\Gamma(\theta_1)$ is the feedthrough of $H_b(\theta_1)$, such that the feedthrough of $H_b$ is being 'replaced' with the true values $\Gamma^0$, and $\Gamma(\theta_1)$ does not appear in the equation.

In (\ref{eq:2claim}) we make no claims on the feedtrough of $H_b$, we have to show that
\begin{equation} \begin{split}
	T'(\theta_1)=T'(\theta_0) \Rightarrow \hspace{4.2cm} &
	\\
	\{
			G(q,\theta^\star),H_a(q,\theta^\star),H_b(q,\theta^\star) - \Gamma(\theta^\star),R(q,\theta^\star)
		&\}
		\\ =
		\{
			G^0(q),H_a^0(q),H_b^0(q) - \Gamma^0,R^0(q).
		&\}
\end{split} \end{equation}
If we consider $\Theta' \in \Theta$ defined by all $\theta$ for which $\Gamma(\theta) = \Gamma^0$,
then using the model set
\[\M' := \left \{ M(\theta) , \; \theta \in \Theta' \right  \} \subseteq \M, \]
we have that $T'(\theta) = T(\theta)$ for all $\theta \in \Theta'$.
This means that we can apply the network identifiability reasoning to this situation.
Since $\M'$ is a subset of $\M$, $\M'$ is globally network identifiable at $M(\theta_0)$ if $\M$ is is globally network identifiable at $M(\theta_0)$.
Using condition \emph{3} we then have that
\begin{equation} \begin{split}
	T'(q,\theta_1) &= T'(q,\theta_0) \\ &\Downarrow \\
	&\left \{
	\begin{array}{ll}
		G(q,\theta_1) &= G^0(q)
		\\
		H_a(q,\theta_1) &= H_a^0(q)
		\\
		H_b(q,\theta_1)-\Gamma(\theta_1) &= H_b^0(q)-\Gamma^0
		\\
		R(q,\theta_1) &= R^0(q).
	\end{array}
	\right \}	
\end{split} \end{equation}
\hfill $\Box$

\section{Proof of Proposition \ref{prop:con}}
The convergence proof in \cite{ljung:99} needs to be adapted slightly in order to prove (\ref{eq:conv_con}).
Under the conditions in part \emph{(1)} the cost function converges
\begin{equation}
	\sup_{\theta \in \Theta}
	\left |
	\frac{1}{N} \sum_{t=1}^N \varepsilon^T_a(t,\theta) Q_a \varepsilon_a(t,\theta)
	-
	\Eb \varepsilon^T_a(t,\theta) Q_a 	\varepsilon_a(t,\theta)
	\right |
	\rightarrow 0
\end{equation}
$w.p. 1$ as $N \rightarrow \infty$.
Similarly the constraint converges
\begin{equation}
	\sup_{\theta \in \Theta}
	\left |
	\frac{1}{N} \sum_{t=1}^N Z^T(t,\theta) Z(t,\theta)
	-
	\Eb Z^T(t,\theta) Z(t,\theta)
	\right |
	\rightarrow 0
\end{equation}
$w.p.$ 1 as $N \rightarrow \infty$.
Since the cost and constraint in (\ref{eq:crit_cons}) both converge (\ref{eq:conv_con}) must hold.

Using the same reasoning as the proof of Proposition \ref{prop:uncon}, $\theta_0$ is a minimum of the cost function, and $\theta_0$ satisfies the constraint.
Now it is shown that $M(q,\theta_0)$ is the only model that is a minimum of the cost function that satisfies the constraint, i.e. $M(q,\theta_0) = M(\theta) \; \forall \; \theta \in \theta^\ast$.

It can be shown  (\citep{ljung:99} proof of Theorem 8.3)   that
\begin{equation} \label{eq:VV1}
	0	=  \Eb \varepsilon_a(t,\theta_0)^T Q_a \varepsilon_a(t,\theta_0) - \Eb \varepsilon_a(t,\theta_1)^T Q_a \varepsilon_a(t,\theta_1)
\end{equation}
if and only if
\begin{equation} \label{eq:VV2}
	0 = \Eb
	(\varepsilon_a(t,\theta_0) - \varepsilon_a(t,\theta_1))^T Q_a (\varepsilon_a(t,\theta_0) - \varepsilon_a(t,\theta_1)).
 \end{equation}

For the constraint we can use the fact that
\begin{equation} \label{eq:zis0}
	Z(t,\theta_0)= \Gamma(\theta_0) \varepsilon_a(t,\theta_0)  - \varepsilon_b(t,\theta_0) =0, \quad \forall t
\end{equation}
up to a possible transient term due to initial conditions that can be neglected in our asymptotic analysis. We can then
rewrite the asymptotic constraint
\begin{equation}
	0 =  \Eb Z^T(\theta_1)Z(\theta_1)
\end{equation}
into the same form as (\ref{eq:VV2})
\begin{equation}
	0 	= \Eb  ( Z(\theta_0)-Z(\theta_1) )^T ( Z(\theta_0)-Z(\theta_1) ).
\end{equation}
Due to condition \emph{(b)} and $Q_a>0$ the predictor filters are identified from the  above two equations, using the definitions of $F_e$ and $F_r$ from the proof of Proposition \ref{prop:uncon}
\begin{align}
	&\left [ \begin{smallmatrix}
	I&0\\ \! \Gamma(\theta_0) \! & -I
	\end{smallmatrix} \right ]
	\left ( F_e(\theta_0) \! + \! \left [ \begin{smallmatrix}
	I\\\Gamma^0
	\end{smallmatrix} \right ] \right )
	\! = \!
	\left [ \begin{smallmatrix}
	I&0\\ \! \Gamma(\theta_1) \! & -I
	\end{smallmatrix} \right ]
	\left ( F_e(\theta_1) \! + \! \left [ \begin{smallmatrix}
	I\\\Gamma^0
	\end{smallmatrix} \right ] \right ),
	\\
	&\left [ \begin{smallmatrix}
	I&0\\\Gamma(\theta_0) & -I
	\end{smallmatrix} \right ]
	F_r(\theta_0)
	=
	\left [ \begin{smallmatrix}
	I&0\\\Gamma(\theta_1) & -I
	\end{smallmatrix} \right ]
	F_r(\theta_1).
\end{align}
In these equations $F_e(\theta_0)=0$ and $F_r(\theta_0)=0$, such that the combination is
\begin{equation}
	\left [ \begin{smallmatrix}
	I & 0 \\ 0 & 0
	\end{smallmatrix} \right ]
	=
	\left [ \begin{smallmatrix}
		I&0\\\Gamma(\theta_1) & -I
	\end{smallmatrix} \right ]
	[
		F_e(\theta_1) + \left [ \begin{smallmatrix}
	I\\\Gamma^0
	\end{smallmatrix} \right ]
	\;
	F_r(\theta_1)
	]	.
\end{equation}
When this equation is pre-multiplied with $(I-G(\theta_1))^{-1} \check H(q,\theta_1) \left [ \begin{smallmatrix}
		I&0\\\Gamma(\theta_1) & -I
	\end{smallmatrix} \right ]$ on both sides, and then $\left [ \begin{smallmatrix}
		0 & (I-G(\theta_1))^{-1} R(q,\theta_1)
	\end{smallmatrix} \right ]$ is added on both sides, it is obtained that
\begin{equation}
	T(q,\theta_0) = T(q,\theta_1),
\end{equation}
By condition \emph{(c)} the model set is globally network identifiable at $\theta_0$ such that
\begin{equation}
	T(\theta_0) = T(\theta_1) \Rightarrow M(\theta_0) = M(\theta_1).
\end{equation}
\hfill $\Box$

\section{Proof of Theorem \ref{thm:ml}}
First the proof of part \emph{1} is given.
The pdf of the innovation $\check e$ is given by 2 equations:
there is the normal distribution of $e=[I \;\; 0] \check e$
\begin{equation}
	f(e) =
	\frac{(2\pi)^{-\frac{p}{2}}}{|\Lambda|^{\frac{1}{2}}}
	\exp
	\left ( -\frac{1}{2} e^T  \Lambda^{-1} e \right ),
\end{equation}
and
\begin{equation}
	\begin{bmatrix}
		\Gamma^0 & -I
	\end{bmatrix}
	\check e =0 \text{ w.p. }1.
\end{equation}
The likelihood for $N$ datapoints is then also given by 2 equations \citep{srivastava2002regression,khatri1968some}
\begin{equation} \label{eq:like_a}
	L_a(\theta) =
	\frac{(2\pi)^{-\frac{pN}{2}}}{|\Lambda(\theta)|^{\frac{N}{2}}}
	\exp
	\left ( -\frac{1}{2} \varepsilon_a^T(t,\theta)  \Lambda^{-1}(\theta) \varepsilon_a(t,\theta) \right ),
\end{equation}
and
\begin{equation} \label{eq:like_b}
	\begin{bmatrix}
		\Gamma(\theta) & -I
	\end{bmatrix}
	\varepsilon(t,\theta) =0  \text{ w.p. }1 \quad \forall t.
\end{equation}
Then taking the natural logarithm results in
\begin{equation} \begin{split}
	\log L_a(\theta) = &c - \frac{N}{2} \log \det \Lambda(\theta)
	\\	&
	- \frac{1}{2} \sum_{t=1}^N \varepsilon_a^T(t,\theta) \Lambda^{-1}(\theta) \varepsilon_a(t,\theta).
\end{split} \end{equation}
$\log L_a(\theta)$ is the criterion to be maximized combined with (\ref{eq:like_b})
\begin{equation}
	\begin{split}
		\theta^{ML}_N = &\arg \max_{\theta} \log L_a(\theta)
		\\
		& \text{subject to } 0 = \varepsilon_b(t,\theta) - \Gamma(\theta) \varepsilon_a(t,\theta) \quad \forall t.
	\end{split}
\end{equation}
Taking the sum of squares for each time $t$ gives the equivalent constraint
\begin{equation}
	  \text{subject to } \frac{1}{N} \sum_{t=1}^N Z^T(t,\theta)Z(t,\theta)=0,
\end{equation}
with $Z$ defined by (\ref{eq:defZ}).

Now part \emph{2} is proven in a similar way as the maximum likelihood proof in \cite{aastrom1980maximum} for full rank noise.
Under the condition that $\Lambda(\theta)$ and $\varepsilon(\theta)$ do not share parameters, the cost function $\log L(\theta)$ is maximized at
\begin{equation} \label{eq:maxL}
	\Lambda(\theta) = \frac{1}{N} \sum_{t=1}^N \varepsilon_a(t,\theta) \varepsilon_a^T(t,\theta)
\end{equation}
In this maximum the constraint of (\ref{eq:ML}) is satisfied.
Then (\ref{eq:maxL}) is substituted into the objective of (\ref{eq:ML}), and added as additional constraint, to obtain (\ref{eq:ML2}).
\hfill $\Box$

\section{Proof of Lemma \ref{lem:rewrite_constrain}}
The constraint is satisfied when $\Pi (\theta - \theta^\ast) =0$ holds.
When substituting (\ref{eq:mapping}) then we have
\begin{equation}
 	\Pi (S \rho + C - \theta^\ast) =0,
\end{equation}
where we have $\Pi S\rho =0$, such that the constraint is independent of $\rho$.
Substituting $C =\Pi^\dagger \Pi \theta^\ast$ then satisfies the equation.
 \hfill $\Box$

\section{Proof of Proposition \ref{prop:crit_rho}}
Proof is by substituting $\theta=S\rho-K^\dagger K^0$ into the CLS (\ref{eq:crit_cons}).
Lemma \ref{lem:rewrite_constrain} shows that this parameter mapping satisfies the constraint for all $\rho$, and thus can be removed.
Equivalence of the cost function is trivial.
\hfill $\Box$

\section{Proof of Proposition \ref{prop:relmap}}
With $P_\theta = \E (\theta^\star - \hat\theta_N)(\theta^\star - \hat\theta_N)^T$
and using the mapping (\ref{eq:mapping}) we get
\begin{equation}
	P_\theta = \E S (\rho^\star - \hat\rho_N)(\rho^\star - \hat\rho_N)^T S^T,
\end{equation}
such that $P_\theta = S P_\rho S^T$.
\hfill $\Box$

\bibliographystyle{plainnat}	
\bibliography{Library_rev}

\end{document}